\documentclass[aps,prc,preprint,superscriptaddress,showpacs,floatfix,nobibnotes]{revtex4}
\usepackage{epsfig,epstopdf,color}

\begin{document}
\begin{center}
{\large\bf A progressive diagonalization scheme for the Rabi
Hamiltonian} \vskip .6cm {\normalsize Feng Pan,$^{a,b}$  Xin
Guan,$^{a}$ Yin Wang,$^{a}$ and J. P. Draayer$^{b}$ } \vskip .2cm
{\small $^{a}$Department of Physics, Liaoning Normal University,
Dalian 116029, P. R. China\vskip .1cm $^{b}$Department of Physics
and Astronomy, Louisiana State University, Baton Rouge, LA
70803-4001, USA}
\end{center}
\vskip .5cm {\bf Abstract:~}\normalsize A diagonalization scheme for
the Rabi Hamiltonian, which describes a qubit interacting with a
single-mode radiation field via a dipole interaction, is proposed.
It is shown that the Rabi Hamiltonian can be solved almost exactly
using a progressive scheme that involves a finite set of one
variable polynomial equations. The scheme is especially efficient
for the lower part of the spectrum. Some low-lying energy levels of
the model with several sets of parameters are calculated and
compared to those provided by the recently proposed generalized
rotating-wave approximation and a full matrix diagonalization.


 \vskip .5cm

It is well known that a qubit (two-level atom) interacting with a
radiation field is a fundamental problem in modern physics,
encountered, for example, in condensed matter, biophysics, and
quantum optics.$^{[1,2]}$ The most common approximation assumes that
the radiation field is quasi-monochromatic, which leads to the
so-called Rabi Hamiltonian.$^{[3,4]}$ Due to its simplicity and
importance, the Rabi Hamiltonian has been studied extensively using
various methods. More complicated extensions to many two-level atoms
interacting with a quasi-mono-chromatic field are described by the
Dicke Model.$^{[5]}$  The Rabi Hamiltonian is also equivalent to
that of the spin-boson model,$^{[2,6]}$ which describes a qubit
interacting with an infinite collection of harmonic oscillators that
model the environment acting as a dissipative bosonic bath. The
latter model has also attracted considerable attention in various
quantum many-body systems due to the rich physics of quantum
criticality and decoherence.$^{[7-9]}$ Though it has long been
conjectured that the Rabi Hamiltonian may be exactly
diagonalizable,$^{[10]}$ with many analytical and numerical studies
carried out for decades,$^{[11-21]}$ exact solutions have not been
presented except for some special cases.$^{[22-31]}$ Due to the
absence of an exact treatment, approximations have been developed.
The rotating wave approximation (RWA),$^{[32]}$ which neglects the
counter-rotating term in the Hamiltonian, a case with the
counter-rotating term, and another case with degenerate level of the
atom are three well-known solvable cases.$^{[30, 31]}$

The Rabi Hamiltonian can be written as$^{[3,4,32]}$

$$\hat{H}_{\rm R}=\omega_{0} a^{\dagger}a+\omega \hat{S}_{0}+g
(a^{\dagger}+a)(\hat{S}_{-}+\hat{S}_{+}),\eqno(1)$$

\noindent where $\omega_{0}$ and $\omega$ are frequencies of a
single-field mode and the level splitting of the two-level atom,
respectively, $g$ is the coupling parameter between the atom and the
field, $a$ and $a^{\dagger}$ are annihilation and creation operators
of the field mode, respectively, satisfying the usual bosonic
commutation relation: $[a,a^{\dagger}]=1$,~$[a,a]=[a^\dagger,
a^\dagger ]=0$, $\hat{S}_{0}$ is the operator of atomic inversion,
and $\hat{S}_{+}$ ($\hat{S}_{-}$) is the atomic raising (lowering)
operator, which satisfy the SU(2) Lie algebraic relations
$[\hat{S}_{0},~\hat{S}_{\pm}]=\pm
\hat{S}_{\pm}$,~$[\hat{S}_{+},~\hat{S}_{-}]=2\hat{S}_{0}$. The
pseudo-spin operator $\hat{S}^{2}=\hat{S}_{+}\hat{S}_{-} +
\hat{S}_{0}(\hat{S}_{0}-1)$ and the parity $\hat{P}
=\exp\left(i\pi(a^{\dagger}a+\hat{S}_{0}+ S)\right)$, where $S=1/2$
is the pseudo spin of the atom, commute with the Hamiltonian (1),
namely $[\hat{S}^{2},\hat{H}_{\rm R}]=0$ and $[\hat{P},\hat{H}_{\rm
R}]=0$.

Let $\vert 0;\mu\rangle$ ($\mu=+$ or $-$) be simultaneously the
boson vacuum and the eigenstate of $\hat{S}_{0}$ satisfying $a\vert
0;\mu\rangle=0, ~\hat{S}_{0}\vert 0;\pm\rangle=\pm{1\over{2}}\vert
0;\pm\rangle$. The Hamiltonian (1) under the basis spanned by $\{
\vert k;\mu\rangle= {1\over{\sqrt{k!}}} a^{\dagger k}\vert
0;\mu\rangle;~~k=0,1,2,\cdots\}$ has the following tridiagonal form:

$$\hat{H}_{\rm R}=\left(\begin{array}{cc}
H^{(+)} & 0 \\
0 & H^{(-)} \\
\end{array}
\right),\eqno(2)$$ of which the order of the Fock states is $\vert
0;+\rangle$, $\vert 1;-\rangle$, $\vert 2;+\rangle$, $\cdots$,
$\cdots$, $\vert 0;-\rangle$, $\vert 1;+\rangle$, $\vert
2;-\rangle$, $\cdots$, where the two tridiagonal infinite
sub-matrices $H^{(+)}$ and $H^{(-)}$ are given by

$$\hat{H}^{(\pm)}=\left(
            \matrix{
              \pm{1\over{2}}\omega & g\sqrt{1} &0~\cdots&&  \cr
              g\sqrt{1} &\omega_{0}\mp {1\over{2}}\omega & g\sqrt{2}&0~\cdots
              \cr
              0& g\sqrt{2} &2\omega_{0}\pm{1\over{2}}\omega& g\sqrt{3}&0\cdots
              \cr
 \vdots&~~~~~~\ddots &~~~~~~\ddots&~~~\ddots&~~~~~~~\ddots\cr}
              ~~\right).\eqno(3)$$

\vskip .3cm \noindent Under the afore arranged order of the basis,
the effective Hamiltonian (3) may be written as

$$\hat{H}^{(\pm)}=(\omega_{0}a^{\dagger}a+g(a+a^{\dagger}))
\pm{1\over{2}}\omega\cos(\pi a^{\dagger}a). \eqno(4)$$ The spin
component $\mu=+$ or $-$ in the Fock states $\vert k;\mu\rangle$
will be omitted in the following for simplicity. The first term in
(4) is easily diagonalizable under the shifted boson states

$$\vert n)=e^{-{1\over{2}}(g/\omega_{0})^2}\sqrt{1\over{n!}}
(a^{\dagger}+g/\omega_{0})^{n}e^{-{g\over{\omega_{0}}}a^{\dagger}}
\vert 0\rangle\eqno(5)$$ with the corresponding eigenvalues
$E_{n}=\omega_{0}n-g^{2}/\omega_{0}$ for $n=0,1,2,\cdots$. The
second term is diagonal in the original boson subspace spanned by
the Fock states $\{\vert k\rangle\}$. Therefore, the Hamiltonian (4)
can also be expressed as

$$\hat{H}^{(\pm)}=\sum_{n=0}^{\infty}E_{n}\vert n)(n\vert\pm
{\omega\over{2}}\mp \omega\sum_{k=0}^{\infty}\vert
2k+1\rangle\langle 2k+1\vert.\eqno(6)$$

Since the structure of the Hamiltonians $\hat{H}^{(+)}$ and
$\hat{H}^{(-)}$ is the same, in the following, we take
$\hat{H}^{(+)}$ to elucidate our diagonalization procedure, and
assume that the parameters $\omega_{0}$, $\omega$, and $g$ are all
non-zero. We adopt the analytical step-by-step diagonalization
procedure proposed in [33]. In the $k$-th step, we take a
diagonalized part of the Hamiltonian $\hat{H}^{(k-1)}$ in the
shifted boson basis (5) and the $k$-th projection in the second term
of (6) with $\hat{H}^{(k)}=\hat{H}^{(k-1)}-\omega\vert
2k+1\rangle\langle 2k+1\vert$ to do the diagonalization. In the
initial $k=0$ step,

$$\hat{H}^{(0)}=\sum_{n=0}^{\infty}E_{n}\vert n)(n\vert+
{\omega\over{2}}-\omega\vert 1\rangle\langle 1\vert,\eqno(7)$$ which
can be diagonalized exactly with the eigenstates

$$\vert \psi^{(0)}_{\tau_{0}}\rangle=
\sqrt{1\over{N^{(0)}_{\tau_{0}}}}\sum_{n=0}^{\infty}
{\alpha^{1}_{n}\over{E_{n}+\omega/2-E^{(0)}_{\tau_{0}}}}\vert
n),\eqno(8)$$ where $N^{(0)}_{\tau_{0}}$ is  the normalization
constant, $\alpha^{2k+1}_{n}$ is the overlap of the original Fock
state $\vert 2k+1\rangle$ with the shifted state $\vert n)$ given by

$$\alpha^{2k+1}_{n}=(n\vert 2k+1\rangle=
\sqrt{(2k+1)!n!e^{-({g\over{\omega_{0}}})^2}}
({g\over{\omega_{0}}})^{n-2k-1}\sum_{q}
{(-)^{q}{(g/{\omega_{0}})}^{2q}\over{(n-2k-1+q)!(2k+1-q)!q!}}.\eqno(9)$$
One can directly check with the eigenequation
$\hat{H}^{(0)}\vert\psi^{(0)}_{\tau_{0}}\rangle
=E^{(0)}_{\tau_{0}}\vert\psi^{(0)}_{\tau_{0}}\rangle$ that (8) is
indeed the eigenstate of $\hat{H}^{(0)}$ with the corresponding
eigenvalue $E^{(0)}_{\tau_{0}}$ if the following equation is
satisfied:

$$1-\omega\sum_{n=0}^{\infty}
{(\alpha^{1}_{n})^2\over{E_{n}+\omega/2-E^{(0)}_{\tau_{0}}}}=0,\eqno(10)$$
where $\tau_{0}$ labels the $\tau_{0}$-th root of Eq. (10). However,
as shown in Fig. 1, the overlaps $\alpha^{2k+1}_{n}$ with fixed $k$
are only non-negligible for a finite number of values $n$ with
$n=m_{1},~m_{1}+1,\cdots,~m_{2}$. As a result, only a finite number
of states $\{\vert n)\}$ correlate among each other in the $k$-th
step,  which, in turn, effectively truncates the infinite sum in
(10) into a finite sum of those terms with non-negligible
$\alpha_{n}^{2k+1}$ for $n=m_{1},~m_{1}+1,\cdots,~m_{2}$. The number
of values $n$ defined by $d_{k}=m_{2}-m_{1}+1$ is called correlation
length. While other states with $n>m_{2}$ or $<m_{1}$ remain
essentially unchanged after the diagonalization. The correlation
length $d_{k}$ increases with increasing $k$ and the coupling
parameter $g/\omega_{0}$. Typically, $d_{k}\sim 10$-$20$ when
$g/\omega_{0}<1$, and $d_{k}\sim 20$-$60$ when $1<g/\omega_{0}<2$.
Furthermore, only those states $\{\vert n)\}$ within the correlated
region will be correlated among each other. In short, the matrix
form of Hamiltonian $\hat{H}^{(k)}$ in the shifted boson basis is
almost block diagonal. As a consequence, Eq. (10) can effectively be
written as

$$1-\omega\sum_{n=0}^{d_{0}-1}
{(\alpha^{1}_{n})^2\over{E_{n}+\omega/2-E^{(0)}_{\tau_{0}}}}=0,\eqno(11)$$
where $m_{1}=0$ and $m_{2}=d_{0}-1$. After the initial step,
$\hat{H}^{(+)}$ given by (6) can be rewritten as

$$\hat{H}^{(+)}=\left(
\sum_{\tau_{0}=0}^{d_{0}-1}E^{(0)}_{\tau_{0}}
\vert\psi^{(0)}_{\tau_{0}}\rangle\langle\psi^{(0)}_{\tau_{0}}\vert+
\sum_{n=d_{0}+1}^{\infty}E_{n}\vert n)( n\vert\right) -
\omega\sum_{k=1}^{\infty}\vert 2k+1\rangle\langle 2k+1\vert,
\eqno(12)$$

\vskip .4cm
\begin{center}
\includegraphics[totalheight=4cm,width=5.51cm]{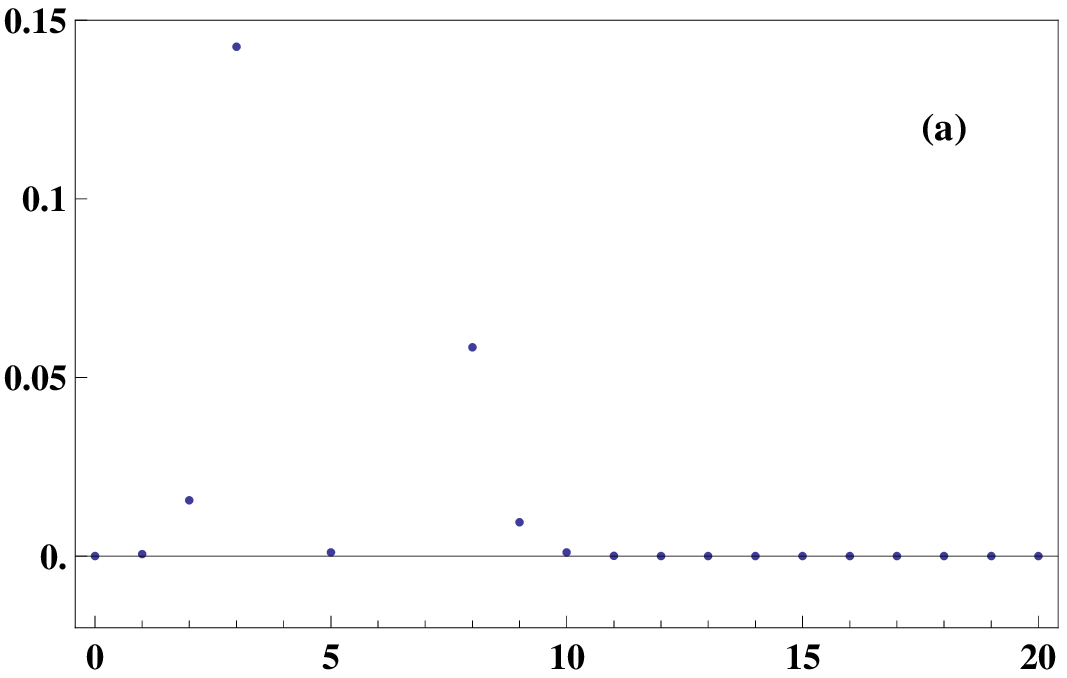}
\includegraphics[totalheight=4cm,width=5.5cm]{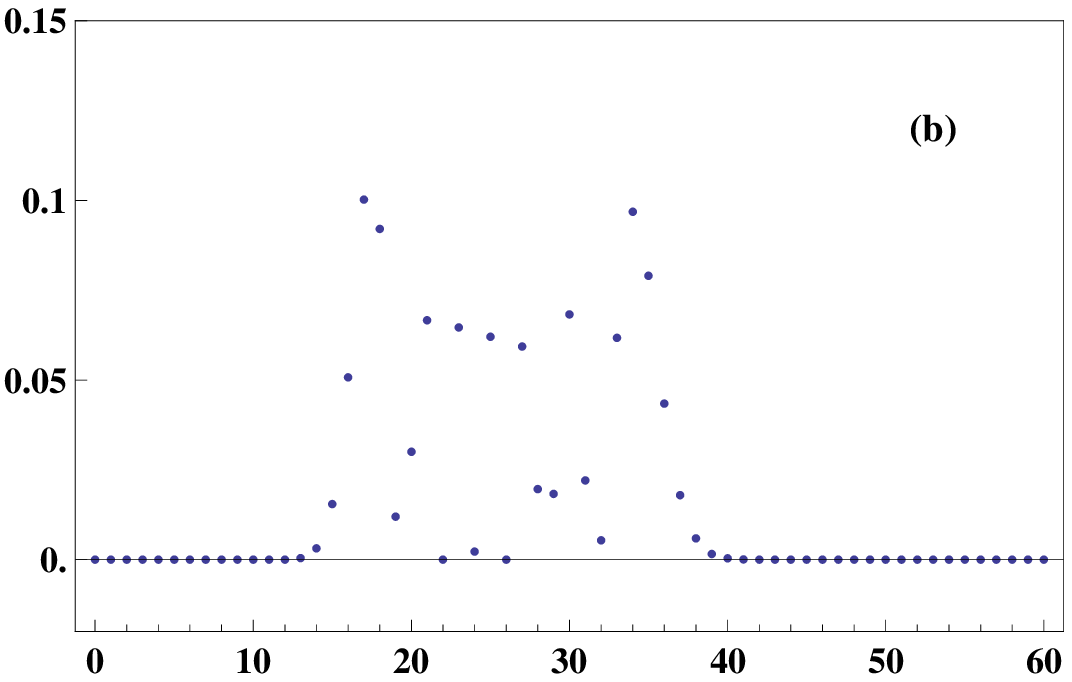}\\
\includegraphics[totalheight=4cm,width=5.5cm]{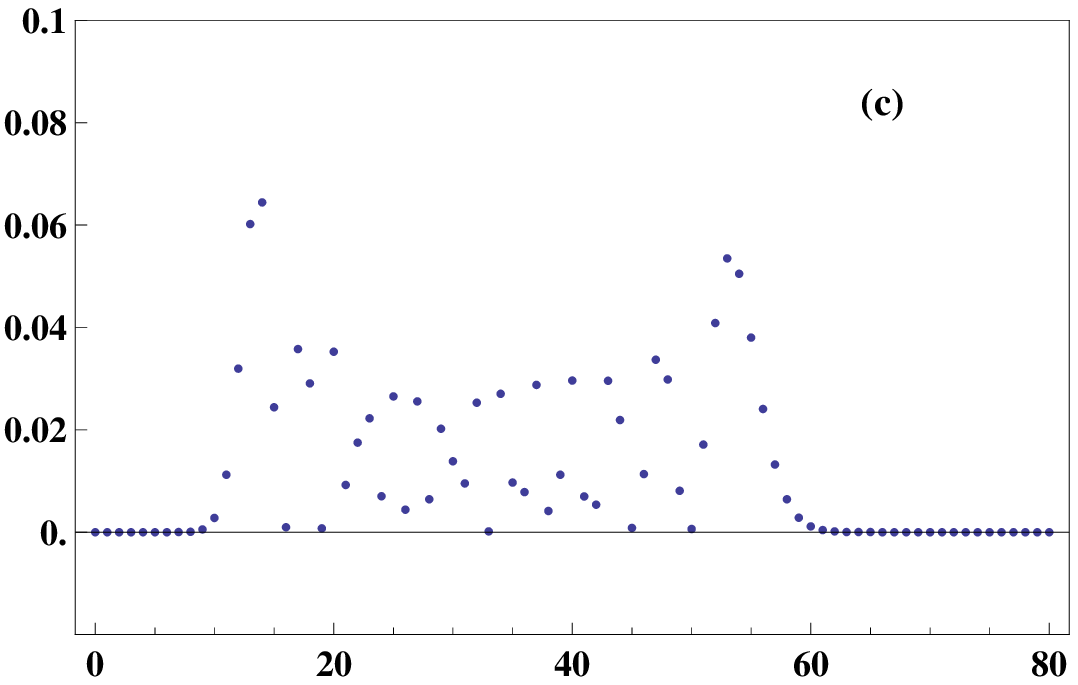}
\includegraphics[totalheight=4cm,width=5.5cm]{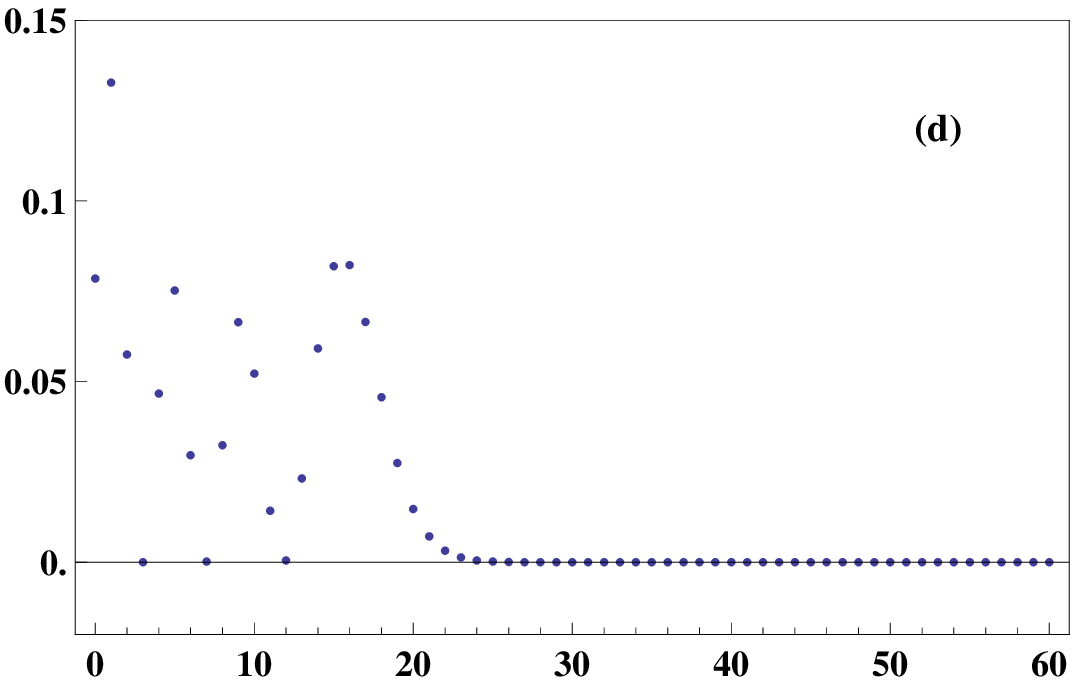}
\end{center}
{\small\bf Fig. 1.} {\small Overlaps $\alpha_{n}^{2k+1}$ as
functions of $n$ for several values of $k$ and the coupling
parameters $g/\omega_{0}$. (a) $k=2$, $g/\omega_{0}=0.5$; (b)
$k=12$, $g/\omega_{0}=1.0$; (c) $k=15$, $g/\omega_{0}=2.0$; (d)
$k=3$, $g/\omega_{0}=2.5$. It can be seen that the corresponding
correlation length is (a) $d_{2}\sim10$, (b) $d_{12}\sim30$, (c)
$d_{15}\sim 50$, and (d) $d_{1}\sim25$, respectively.} \vskip .4cm

\noindent where
$$\vert \psi^{(0)}_{\tau_{0}}\rangle=
\sqrt{1\over{N^{(0)}_{\tau_{0}}}}\sum_{n=0}^{d_{0}-1}
{\alpha^{1}_{n}\over{E_{n}+\omega/2-E^{(0)}_{\tau_{0}}}}\vert
n).\eqno(13)$$

Next, using a similar procedure, the first term and the $k=1$
projection in the second term of (12) given by

$$\hat{H}^{(1)}=\sum_{\tau_{0}=0}^{d_{0}-1}E^{(0)}_{\tau_{0}}
\vert\psi^{(0)}_{\tau_{0}}\rangle\langle\psi^{(0)}_{\tau_{0}}\vert+
\sum_{n=d_{0}}^{\infty}E_{n}\vert n)( n\vert - \omega\vert
3\rangle\langle3\vert  \eqno(14)$$ needs to be diagonalized. If the
correlation length $d_{1}>d_{0}$, the correlated eigenstates may be
written as

$$\vert \psi^{(1)}_{\tau_{1}}\rangle=
\sqrt{1\over{N^{(1)}_{\tau_{1}}}} \left(\sum_{\tau_{0}=l}^{d_{0}-1}
{\langle\psi^{(0)}_{\tau_{0}}\vert
3\rangle\over{E^{(0)}_{\tau_{0}}-E^{(1)}_{\tau_{1}}}}
\vert\psi^{(0)}_{\tau_{0}}\rangle+ \sum_{n=d_{0}}^{d_{1}-1}
{\alpha^{3}_{n}\over{E_{n}+\omega/2-E^{(1)}_{\tau_{1}}}}\vert n)
\right),\eqno(15)$$ where it is assumed that
$\langle\psi^{(0)}_{\tau_{0}}\vert 3\rangle\sim0$ when $\tau_{0}<l$,
and

$$\langle\psi^{(0)}_{\tau_{0}}\vert
3\rangle=\sqrt{1\over{N^{(0)}_{\tau_{0}}}}\sum_{n=0}^{d_{0}-1}
{\alpha^{1}_{n}\alpha^{3}_{n}\over{E_{n}
+\omega/2-E^{(0)}_{\tau_{0}}}}\eqno(16)$$ according to (13), while
the uncorrelated eigenstates satisfy
$\vert\psi^{(1)}_{\tau}\rangle=\vert\psi^{(0)}_{\tau}\rangle$ with
the corresponding eigenvalue $E^{(0)}_{\tau}$ for $\tau<l$, and
$\vert\psi^{(1)}_{n}\rangle=\vert n)$ with the corresponding
eigenvalue $E_{n}$ for $n\geq d_{1}$. Using (14) and (15) to solve
the eigenequation $\hat{H}^{(1)}\vert\psi^{(1)}_{\tau_{1}}\rangle
=E^{(1)}_{\tau_{1}}\vert\psi^{(1)}_{\tau_{1}}\rangle$, one can prove
that (15) is indeed the eigenstate of $\hat{H}^{(1)}$ if the
corresponding eigenvalue $E^{(1)}_{\tau_{1}}$ is the $\tau_{1}$-th
root of the following equation:

$$1-\omega\left(\sum_{\tau_{0}=l}^{d_{0}-1}
{(\langle\psi^{(0)}_{\tau_{0}}\vert
3\rangle)^{2}\over{E^{(0)}_{\tau_{0}}-E^{(1)}_{\tau_{1}}}}
+\sum_{n=d_{0}}^{d_{1}-1}
{(\alpha^{3}_{n})^2\over{E_{n}+\omega/2-E^{(1)}_{\tau_{1}}}}\right)
=0.\eqno(17)$$

According to this procedure, once eigenvectors
$\vert\psi^{(k)}_{\tau_{k}}\rangle$ and the corresponding
eigenvalues $E^{(k)}_{\tau_{k}}$ are known from the $k$-th step, the
$k+1$-th step results are given by

$$\vert\psi^{(k+1)}_{\tau_{k+1}}\rangle=
\sqrt{1\over{N^{(k+1)}_{\tau_{k+1}}}}\sum_{\tau_{k}=m_{1}}^{m_{2}}
{\langle\psi^{(k)}_{\tau_{k}}\vert
2k+1\rangle\over{E^{(k)}_{\tau_{k}}-E^{(k+1)}_{\tau_{k+1}}}}
\vert\psi^{(k)}_{\tau_{k}}\rangle\eqno(18)$$ for
$\tau_{k+1}=m_{1},m_{1}+1,\cdots,m_{2}$ and $d_{k+1}=m_{2}-m_{1}+1$,
in which $E^{(k+1)}_{\tau_{k+1}}$ should satisfy

$$1-\omega\sum_{\tau_{k}=m_{1}}^{m_{2}}
{(\langle\psi^{(k)}_{\tau_{k}}\vert
2k+1\rangle)^{2}\over{E^{(k)}_{\tau_{k}}-E^{(k+1)}_{\tau_{k+1}}}}=0.
\eqno(19)$$ While
$\vert\psi^{(k+1)}_{\tau_{k+1}}\rangle=\vert\psi^{(k)}_{\tau_{k+1}}\rangle$
with the corresponding eigenvalue
$E^{(k+1)}_{\tau_{k+1}}=E^{(k)}_{\tau_{k+1}}$  for
$\tau_{k+1}=0,1,\cdots,m_{1}-1$, and
$\tau_{k+1}=m_{2}+1,m_{2}+2,\cdots$. In fact, with increasing $k$,
the region of the correlation among $\{\vert n)\}$ moves to larger
values of $n$, and keeps lower part of the levels with small $n$
less affected. Therefore, after a number of steps, several of the
low-lying eigenstates of the Hamiltonian (6) and the corresponding
eigenvalues can be obtained. The results are almost exact because
the overlaps $\langle\psi^{(k)}_{\tau_{k}}\vert 2k+1\rangle$ are
small enough to be neglected when $\tau_{k}$ satisfies
$\langle\psi^{(k)}_{\tau_{k}}\vert 2k+1\rangle\sim0$, which is
similar to the situations shown in Fig. 1 for
$\alpha_{n}^{2k+1}=(n\vert 2k+1\rangle$ because the overlaps
$\langle\psi^{(k)}_{\tau_{k}}\vert 2k+1\rangle$ is a linear function
of $\alpha_{n}^{2k+1}=(n\vert 2k+1\rangle$.

\begin{table} \caption{\small The first few low-lying
energy levels (ordered from the lowest) determined by
the Rabi Hamiltonian (1) obtained from the
Progressive Diagonalization Scheme (PDS) are compared to those
calculated using the GRWA and results from a Full Matrix
Diagonalization (FMD), which is done by diagonalizing
the infinite tridiagonal matrix (3) in a truncated Hilbert subspace.}
\vskip .3cm
\begin{tabular}{ccccccccccccccccc}
\hline \hline &~~$g/\omega_{0}=0.5,~\omega_{0}=\omega$
&$g/\omega_{0}=1.0,~\omega_{0}=\omega$
&~~$g/\omega_{0}=2.0,~\omega_{0}=0.75\omega$\\
\hline
&~PDS~~~~~~~GRWA~~~~~~FMD   &~~~~PDS~~~~~~GRWA~~~~FMD~~  &~~~PDS~~~~~GRWA~~~~FMD\\
\hline
&-0.6332~~~~~-0.5532~~~~~-0.6333~&-1.1479~~~~-1.0677~~~-1.1479&~~-3.0225~~~-3.0002~~~-3.0225\\
&-0.1200~~~~~-0.0602~~~~~-0.1200~&-1.0100~~~~-0.9936~~~-1.0102&~~-3.0221~~~-2.9999~~~-3.0222\\
&~0.6953~~~~~~0.7401~~~~~~0.6954~&-0.2317~~~~-0.2110~~~-0.2317&~~-2.2793~~~-2.2546~~~-2.2794\\
&~0.8253~~~~~~0.8635~~~~~~0.8253~&~0.1334~~~~~~0.2642~~~~0.1334&~~-2.2738~~~-2.2474~~~-2.2739\\
&~1.5870~~~~~~1.5977~~~~~~1.5870~&~0.9270~~~~~~0.9403~~~~0.9270&~~-1.5524~~~-1.5142~~~-1.5525\\
&~1.9355~~~~~~1.9115~~~~~~1.9355~&~1.1048~~~~~~1.0539~~~~1.1048&~~-1.5172~~~-1.4966~~~-1.5172\\
&~2.5485~~~~~~2.5375~~~~~~2.5485~&~1.8428~~~~~~1.8559~~~~1.8429&~~-0.8554~~~-0.8385~~~-0.8554\\
&~2.9477~~~~~~2.9528~~~~~~2.9478~&~2.1436~~~~~~2.1269~~~~2.1437&~~~-0.7376~~-0.6792~~~-0.7376\\
\hline \hline\\
\end{tabular}
\end{table}

Since the Rabi Hamiltonian can not be solved exactly except
the case with the RWA$^{[32]}$ and other two special cases studied in [30-31].
As shown in [20], the Generalized Rotating-Wave Approximation (GRWA)
is significantly more accurate than the
RWA. To demonstrate our Progressive Diagonalization Scheme (PDS), a few
low-lying energy levels with three sets of model parameters are
calculated and compared to those calculated using the GRWA and
a Full Matrix Diagonalization (FMD), which is done by diagonalizing
the infinite tridiagonal matrix (3) in a truncated Hilbert subspace.
The dimension of the truncated subspace, with which the first few
low-lying energies keep unchanged with further increasing of the
dimension, depends on the coupling
parameter $g/\omega_{0}$. For $g/\omega_{0}=0.5${-}$2.0$,
the dimension of the truncated subspace is $10^3$ ${-}$ $10^5$,
with which the FMD can be preformed on a PC.
When $g/\omega_{0}>2.0$, in order
to get accurate results, the FMD needs more computer memory
to store matrix elements and many hours of CPU time to do
the diagonalization. Therefore, we only compare the PDS with
the GRWA and the FMD for $g/\omega_{0}\leq 2.0$.
The results are shown in Table 1.
The PDS results were obtained with less then $4$ steps for
$g/\omega_{0}=0.5$, $14$ steps for $g/\omega_{0}=1.0$, and less than
$30$ steps for $g/\omega_{0}=2.0$, which are exactly the same as the
corresponding results obtained from a full matrix diagonalization.
In each step, only a single variable finite
order polynomial equation needs to be solved.
The order of the polynomial equations used in the PDS
depends on the correlation length $d_{k}$. In the cases
shown in Table 1, the maximal $d_{k}$ is around $60$
when  $g/\omega_{0}=2.0$. Such polynomial equations
can easily be solved by using Mathematica with a few seconds
of CPU time. Though the PDS for  $g/\omega_{0}=2.0$ needs
about $30$ steps in order to get accurate results, the total CPU
time needed is much less than that of the FMD. Besides the FMD
needs more memory, the CPU time needed is $10${-}$20$ times
more than the total CPU time needed for the PDS.
The PDS results are certainly better than those obtained from the
GRWA, which nonetheless has the advantage of being a simpler to
implement.

In conclusion, a diagonalization scheme of the Rabi Hamiltonian in
the shifted boson basis is proposed based on the step-by step
diagonalization method, from which lower part of eigenstates and the
corresponding eigenvalues can be obtained progressively within a
finite number of steps. In each step, only a single variable finite
order polynomial equation needs to be solved due to the fact that
only a finite number of the shifted boson states correlate among
each other. Such type of polynomial equations also appear
in the Tamm-Dancoff and random phase approximations with
separable potentials in nuclear physics.$^{[34]}$
The convergence is related to the number of steps that
are needed, which in turn depends on the coupling parameter
$g/\omega_{0}$ and and position of the energy levels to be
calculated. Only a few steps are needed for $g/\omega_{0}<1$, while
$10$-$30$ steps are needed in order to get accurate results for the
low-lying part of the spectrum for $1< g/\omega_{0}\leq 2$.
More steps are needed for higher excited states.
Anyway, the total CPU time needed is much less than that of
the truncated matrix diagonalization for the
low-lying part of the spectrum. As shown in [20], the
coupling $g/\omega_{0}$ is often small with $g/\omega_{0}<1$ in
quantum optics, microwave resonator, superconducting LC circuits,
but $g/\omega_{0}\sim 1$ for nano-mechanical resonator coupled to a
charged qubit. Higher values of $g/\omega_{0}$ may also be possible
for spin-boson models of other systems. Since the PDS scheme yields
nearly exact results that seem to be independent of the model
parameters $\omega_{0}$, $g$, and $\omega$, and the single variable
finite order polynomial equation involved in each step is much
more easily to be solved than the truncated matrix diagonalization,
it should be a useful scheme for studying dynamics in the
spin-boson model for strong coupling cases, especially in
studying quantum critical phenomena and decoherence, where
accurate solutions to the problem is essential.
Related work is in progress.

\vskip .2cm  Support from the U.S. National Science Foundation
(PHY-0500291 \& OCI-0904874), the Southeastern Universities Research Association, the
Natural Science Foundation of China (10775064), the Liaoning
Education Department Fund (2007R28), and the LSU--LNNU joint
research program (9961) is acknowledged.

\vskip .5cm

\def\HT{\bf\relax}
\def\REF#1{\small\par\hangindent\parindent\indent\llap{#1\enspace}\ignorespaces}

\section*{References}

\noindent\REF{[1]} L. Allen and J. H. Eberly, Optical Resonance and
Two-Level Atoms (Wiley, New York, 1975).

\REF{[2]} A. J. Leggett, S. Chakravarty, A. T. Dorsey, M. P. A.
Fisher, A. Garg, and W. Zwerger, Rev. Mod. Phys. {\bf 59}, 1 (1987).

\noindent{[3]} I. I. Rabi, Phys. Rev. {\bf 49}, 324 (1926).

\noindent{[4]} I. I. Rabi, Phys. Rev {\bf 51}, 652 (1937).

\REF{[5]} R. Dicke, Phys. Rev. {\bf 93}, 99 (1954).

\REF{[6]} M. Blume, V. J. Emery, and A. Luther, Phys. Rev. Lett.
{\bf 25}, 450 (1970).

\REF{[7]} K. L. Hur, P. D. Beaupr\'{e}, and W. Hofstetter, Phys.
Rev. Lett. {\bf 99}, 126801(2007).

\REF{[8]} A. Kopp and K. L. Hur, Phys. Rev. Lett. {\bf 98},
220401(2007).

\REF{[9]} U. Weiss, Quantum Dissipative Systems (World Scientific,
Singapore, 1993).

\REF{[10]} H. G. Reik and M. Doucha, Phys. Rev. Lett. {\bf 57}, 787
(1986).

\REF{[11]} T. Yabuzaki, S. Nakayama, Y. Murakami, and T. Ogawa,
Phys. Rev. A {\bf 10}, 1955 (1974).

\REF{[12]} M. Kus, Phys. Rev. Lett. {\bf 54}, 1343 (1985).

\REF{[13]} L. M\"{u}ller, J. Stolze, H. Leschke, and P. Nagel, Phys.
Rev. A {\bf 44}, 1022 (1991).

\REF{[14]} L. Bonci, R. Roncaglia, B. J. West, and P. Grigolini,
Phys. Rev. Lett. {\bf 67}, 2593 (1991).

\REF{[15]} L. Bonci and P. Grigolini, Phys. Rev.  A {\bf 46}, 4445
(1992).

\REF{[16]} R. F. Bishop, N. J. Davidson, R. M. Quick, and D. M. van
der Walt, Phys. Rev. A {\bf 54}, R4657 (1996).

\REF{[17]} A. Pereverzev and E. R. Bittner, Phys. Chem. Chem. Phys.
{\bf 8}, 1378 (2006).

\REF{[18]} V. Fessatidis, J. D. Mancini, S. P. Bowenb, Phys. Lett. A
{\bf 297}, 100 (2002).

\REF{[19]} R. Graham and M. H\"{o}hnerbach, Z. Phys. B {\bf 57}, 233
(1984).

\REF{[20]} E. K. Irish, Phys. Rev. Lett. {\bf 99}, 173601 (2007);
Phys, Rev. Lett. {\bf 99}, 259901 (E) (2007).

\REF{[21]} T. Liu, K. L. Wang and M. Feng, EPL {\bf 86}, 54003
(2009).

\REF{[22]} V. Loorits, J. Phys C {\bf 16}, L711 (1983).

\REF{[23]} M. Kus and M. Lewenstein, J. Phys. A {\bf 19}, 305
(1986).

\REF{[24]} R. Ko\c{c}, H\"{u}t\"{u}nc\"{u}ler, and M. Koca, J. Phys.
A {\bf 35}, 9425 (2002).

\REF{[25]} R. F. Bishop, N. J. Davidson, R. M. Quick, and D. M. van
der Walt, Phys. Rev. A {\bf 54}, R4657 (1996).

\REF{[26]} M. Szopa, G. Mys, A. Ceulemans, J. Math. Phys. {\bf 37},
5402 (1996).

\REF{[27]} C. Emary and R. F. Bishop, J. Math. Phys. {\bf 43}, 3916
(2002).

\REF{[28]} S. N. Dolya and O. B. Zaslavskii, J. Phys. A {\bf 33},
L369 (2000).

\REF{[29]} S. N. Dolya and O. B. Zaslavskii, J. Phys. A {\bf 34}
5955 (2001).

\REF{[30]} Feng Pan, Y.-K. Yao, M.-X. Xie, W.-J. Han, and J. P.
Draayer, Commun. Theor. Phys. {\bf 48}, 53 (2007).

\REF{[31]} E. K. Irish, J. Gea-Banacloche, I. Martin, and K. C.
Schwab, Phys. Rev. B {\bf 72}, 195410 (2005).

\REF{[32]} E. T. Jaynes and F. W. Cummings, Proc. IEEE {\bf 51}, 89
(1963).

\REF{[33]} Feng Pan, Ming-Xia Xie, Xin Guan, Lian-Rong Dai, and J.
P. Draayer, Phys. Rev. C {\bf 80}, 044306 (2009).

\REF{[34]} P. Ring and P. Schuck, The Nuclear Many-Body Problem
(Springer-Verlag, Berlin, 1980).

\end{document}